\documentclass[aps,10pt,pra,twocolumn,amssymb,amsmath,amsfonts,showpacs,floatfix,citeautoscript,preprintnumbers,footinbib,groupedaddress,a4paper]{revtex4-1}
\usepackage{times}
\usepackage{graphicx}
\usepackage[latin1]{inputenc}

\newcommand{\mymat}[1]{\big[{#1}\big]}
\newcommand{\ket}[1]{\vert{#1}\rangle}
\newcommand{\bra}[1]{\langle{#1}\vert}
\newcommand{\braket}[2]{\langle{#1}|{#2}\rangle}
\newcommand{\moy}[1]{\langle{#1}\rangle}
\newcommand{\elem}[3]{\langle{#1}\vert{#2}\vert{#3}\rangle}
\newcommand{\modul}[1]{\vert{#1}\vert}
\newcommand{\ups}{\uparrow}
\newcommand{\downs}{\downarrow}
\newcommand{\Ham}{\mathcal{H}}
\newcommand{\MGstate}{\includegraphics[width=3.2cm,clip]{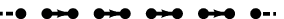}}
\newcommand{\dimer}{\includegraphics[width=4mm,clip]{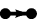}}
\newcommand{\MGstateEx}{\includegraphics[width=3.2cm,clip]{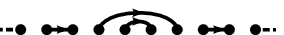}}
\newcommand{\MGstateLongEx}{\includegraphics[width=3.2cm,clip]{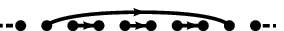}}
\newcommand{\MGstateExCross}{\includegraphics[width=3.2cm,clip]{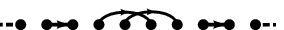}}
\newcommand{\spinon}{\includegraphics[width=2.2cm,clip]{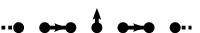}}
\newcommand{\spinonHopeRight}{\includegraphics[width=2.2cm,clip]{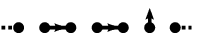}}

\newcommand{\spinonExcited}{\raisebox{-1mm}{\includegraphics[width=2.2cm,clip]{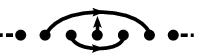}}}
\newcommand{\spinonExcitedRoofRight}{\raisebox{-1mm}{\includegraphics[width=2.2cm,clip]{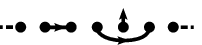}}}

\newcommand{\LongDimerExcitedCenter}{\includegraphics[width=4.2cm,clip]{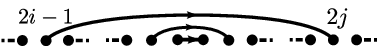}}
\newcommand{\LongDimerExcitedRight}{\includegraphics[width=4.5cm,clip]{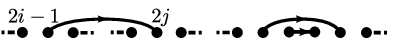}}
\newcommand{\LongDimerExcitedLeft}{\includegraphics[width=4.5cm,clip]{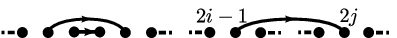}}
\newcommand{\LongDimerRoofLeft}{\raisebox{-3mm}{\includegraphics[width=4.2cm,clip]{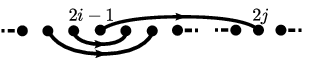}}}
\newcommand{\LongDimerRoofRight}{\raisebox{-3mm}{\includegraphics[width=4.2cm,clip]{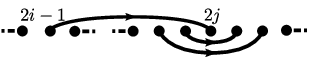}}}
\newcommand{\energy}{E} 

\begin{document}

\def\lptms{LPTMS, Univ.~Paris-Sud, CNRS, UMR8626, F-91405 Orsay, France.}

\title{Localization of spinons in random Majumdar-Ghosh chains}
\author{Arthur Lavar\'elo}
\author{Guillaume Roux}
\affiliation{\lptms}

\date{\today}

\begin{abstract}
  We study the effect of disorder on frustrated dimerized spin-$1/2$
  chains at the Majumdar-Ghosh point. Using variational methods and
  density-matrix renormalization group approaches, we identify two
  localization mechanisms for spinons which are the deconfined
  fractional elementary excitations of these chains. The first one
  belongs to the Anderson localization class and dominates at the
  random Majumdar-Ghosh (RMG) point. There, spinons are almost
  independent, remain gapped, and localize in Lifshitz states whose
  localization length is analytically obtained. The RMG point then
  displays a quantum phase transition to phase of localized spinons at
  large disorder. The other mechanism is a random confinement
  mechanism which induces an effective interaction between spinons and
  brings the chain into a gapless and partially polarized phase for
  arbitrarily small disorder.
\end{abstract}


\maketitle

Spinons are fractional excitations corresponding to half of a spin
excitation in quantum magnets. They typically appear in understanding
the excitation spectrum of one-dimensional systems such as the
frustrated $J_1-J_2$ Heisenberg chain. This model possesses an exact
ground-state at the Majumdar-Ghosh (MG) point $J_1 =
2J_2$~\cite{Majumdar1969} which is the prototype of a valence bond
solid (VBS) state and for which a variational approach describes well
elementary excitations~\cite{Shastry1981}. Further, spinons play a
crucial role in unconventional two-dimensional phase transitions in
which they could be deconfined~\cite{Senthil2004}. Investigating the
effect of disorder on their dynamics is all the more essential, since
randomness is inherent to experimental samples. Possible strategies to
study random quantum magnets are bosonization~\cite{Giamarchi2004},
provided the disorder is small, or real-space renormalization group
(RSRG)~\cite{Ma1979}, rather suited for the strong disorder
regime. The latter is asymptotically exact in the case of an
infinite-disorder fixed point~\cite{Fisher1994}, but when it converges
to a finite-disorder fixed point, its outcome can be questioned at
small disorder. Numerical approaches are challenging due to strong
finite-size effects from rare events~\cite{Igloi2000} and the
interplay between frustration and disorder cannot be addressed using
the powerful quantum Monte-Carlo method because of the sign
problem. Lastly, most studies on random magnets focus on the
ground-state while little is known about the fate of elementary
excitations. So far, it has been conjectured~\cite{Yang1996} that the
gap of frustrated dimerized chains is broken by a domain formation
mechanism similar to the one suggested for Mott
phases~\cite{Shankar1990}. Later, RSRG studies~\cite{Hoyos2004} found
that it would belong to the class of the large-spin
phase~\cite{Westerberg1997}.

In this Letter, two localization mechanisms at play in random
frustrated dimerized chains are unveiled using a variational approach
supported by density-matrix renormalization group (DMRG)
calculations~\cite{White1992}. They provide both quantitative
predictions and an intuitive picture of the physics. The first
mechanism belongs to the Anderson class and governs the dynamics of a
spinon at the random Majumdar-Ghosh (RMG) point which generalizes the
MG condition in the presence of random bonds. Increasing disorder at
the RMG point induces a transition to a paramagnetic phase of
localized spinons. The second one is a random confinement which
generates an effective interaction between spinons which stabilizes
the formation of domains and breaks the spin gap.

\begin{figure*}[t]
\centering
\includegraphics[width=0.9\textwidth,clip]{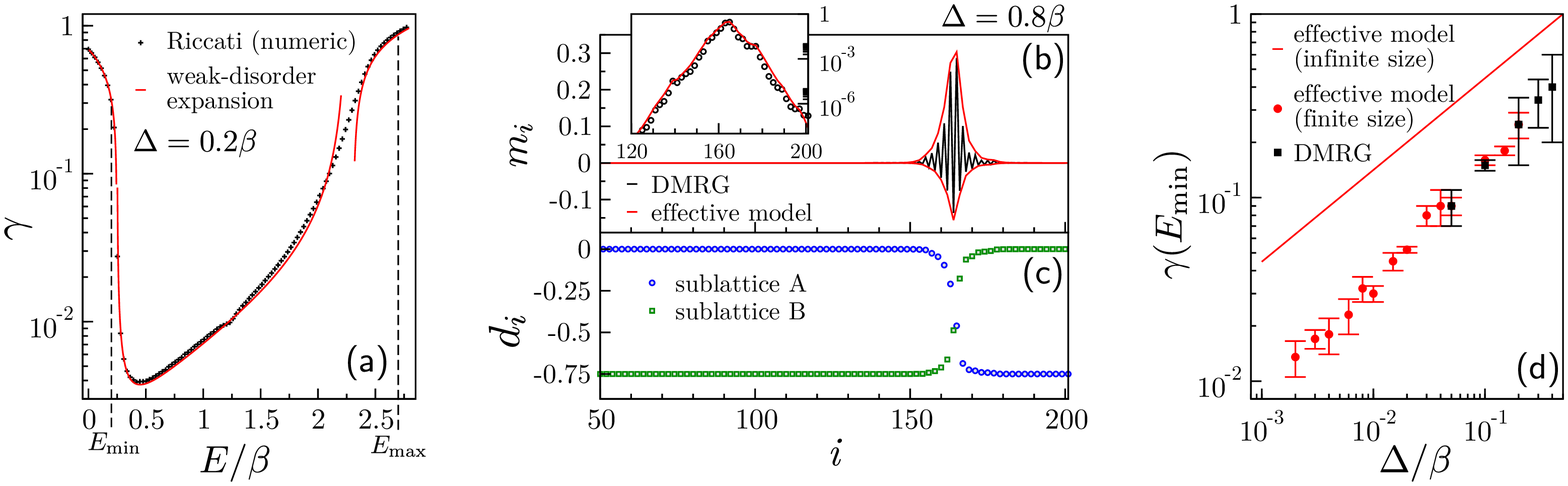}
\caption{(color online) \emph{At the RMG point} :
(a) Lyapunov exponent $\gamma$ vs energy $\energy$ of the effective model. 
(b) magnetization profile from DMRG in sector $S^z=1/2$. Inset: log plot.
(c) dimerization profile showing the MG domains.
(d) finite size effects on the Lyapunov exponent.
}
\label{fig:localization}
\end{figure*}

\emph{Model} -- We consider a frustrated dimerized spin-$1/2$ chain
with random nearest-neighbor couplings $\alpha_i$ and next-nearest
neighbor couplings $\beta_i$:
\begin{equation}
\Ham = \sum_{i} \alpha_i \mathbf{S}_i\cdot\mathbf{S}_{i+1} + \beta_i \mathbf{S}_{i-1}\cdot\mathbf{S}_{i+1}\;,
\label{eq:hamiltonian}
\end{equation}
where $\mathbf{S}_i$ are spin-$1/2$ operators. In the following, the
average couplings are written $\overline{\alpha_i}=\alpha$ and
$\overline{\beta_i} = \beta$ with $\alpha = 2\beta$ to start from the
usual MG point. When applying $\Ham$ on MG states $\ket{\text{MG}} =
\ket{\MGstate}$ with dimers $\ket{\dimer} =
\frac{1}{\sqrt{2}}[\ket{\ups\downs}-\ket{\downs\ups}]$, it can be
shown~\cite{Yang1996,supplementary} that the latter remain degenerate
eigenstates of the Hamiltonian provided
\begin{equation}
\alpha_{i} = \beta_{i} + \beta_{i+1}\;,
\label{eq:MGcondition}
\end{equation}
which we call the random Majumdar-Ghosh (RMG) point. It imposes a
local correlation between the random couplings. The two localization
mechanisms stem from the splitting of the Hamiltonian into $\Ham =
\Ham_{\text{RMG}} + \Ham_{\text{dim}}$, where $\Ham_{\text{RMG}}$
follows \eqref{eq:MGcondition} and $\Ham_{\text{dim}}$ is the
remaining part (see below).

\emph{Localization at the RMG point} -- An effective model for the
dynamics of a spinon is obtained assuming \eqref{eq:MGcondition} by
considering an open chain with an odd number of sites and spinon
states $\ket{j}=\ket{\spinon}$ with a free spin at site $2j+1$
separating two MG domains. By projecting~\cite{supplementary}
$\Ham_{\text{RMG}}$ orthogonally onto the free family of
non-orthogonal states $\{\ket{j}\}$, the effective spinon Hamiltonian
reads
\begin{equation}
\left(\widetilde{\Ham}_{\text{RMG}}-E_{\text{MG}}\right)\ket{j} = \frac{\beta_{2j+1}}{2}\Big(\ket{j-1}+\frac{5}{2}\ket{j}+\ket{j+1}\Big)\;,
\label{eq:spinon-Ham}
\end{equation}
where $E_{\text{MG}} = -\frac 3 4 \sum_i\beta_i$ is the MG state
energy extrapolated to odd sizes. Consequently, the motion of the
spinon, taking place either on odd or even site sublattices, obeys a
special kind of Anderson Hamiltonian expected to induce localization.
The matrix of $\widetilde{\Ham}_{\text{RMG}}$ is non-hermitian because
$\{\ket{j}\}$ is not orthogonal but is similar to an hermitian
matrix~\cite{supplementary}. Yet, tridiagonal form is well suited to
the Dyson-Schmidt method~\cite{Dyson1953, Luck1992}.  By writing
$\ket{\psi} = \sum_j \psi_j\ket{j}$ the spinon variational
wave-function and $\tilde{\beta}_j \equiv \beta_{2j+1}$, we introduce
the Riccati variables $R_j = \psi_{j+1}/\psi_{j}$ to rewrite
Schr\"odinger's equation as
\begin{equation}
R_j + 5/2 + 1/R_{j-1} = 2\energy/\tilde{\beta}_j \;,
\label{eq:Riccati}
\end{equation}
for a given spinon energy $\energy$. The integrated density of states
$N(\energy)$ and Lyapunov exponent $\gamma(\energy)$ of a single
spinon excitation are obtained by extending the energy to the complex
plane. Assuming that the probability density of the Riccati variables
converges toward an invariant distribution of measure $dW(R)$ as
$j\rightarrow\infty$, the characteristic function
\begin{equation}
\Omega(z) = \int dW(R) \ln{R}
\label{eq:characteristic}
\end{equation}
is such that $\Omega(\energy+i0^+) = \gamma(\energy) +
i\pi(1-N(\energy))$. These quantities can be obtained either
numerically or analytically from a weak-disorder
expansion~\cite{Nieuwenhuizen1982} as described below.  In the
non-disordered case, the spinon dispersion relation is
$\energy(k)=\beta(5/4 + \cos{k})$ with $k$ the momentum.  By
introducing the variables
\begin{equation*}
y_j = \bigg[1+ \frac{y_{j-1} - 1}{1+g_j(y_{j-1}-1)}\bigg] e^{-2ik},\; 
g_j = \frac{\energy(k)}{i\sin{k}}\bigg(\frac 1 \beta - \frac 1 {\tilde{\beta}_j}\bigg),
\label{eq:weak-coupling}
\end{equation*}
it can be shown that the first term in the expansion in the first
moment of the $g$-distribution gives $\Omega \simeq ik -
\overline{g^2}/2$. Specializing to the case of a uniform distribution
over $[\beta-\Delta,\beta+\Delta]$, the explicit calculation for
energies $\energy < \beta/4$ gives~\cite{supplementary}
\begin{equation} 
\gamma(\energy) = \text{arcosh}{\bigg[\frac 5 4 - \frac E \beta\bigg]} - \frac{1}{6}\frac{\energy^2}{(E-\frac{5}{4}\beta)^2-\beta^2}\bigg(\frac{\Delta}{\beta}\bigg)^2.
\end{equation}
The result is compared to numerics on Fig~\ref{fig:localization}(a).
In particular, we obtain that the RMG spinon localization length
$\xi_{\text{RMG}}=1/\gamma(E_{\text{min}})$ in the state with the
lowest energy $E_{\text{min}} = \beta_{\text{min}}/4$ (where
$\beta_{\text{min}} = \min\tilde{\beta}_j$), scales as:
\begin{equation}
\xi_{\text{RMG}} \simeq \sqrt{2\beta/\Delta}\;.
\label{eq:xi-weak-coupling}
\end{equation}
Notice that $\xi_{\text{RMG}}$ cannot be captured by RSRG and is
not related to the spin correlation length of a MG state.

Another important outcome of the effective model is that it provides
hints on the finite-size effects on the spinon energies, with
consequences on the spin gap and the localization length. The lowest
energy spinon states correspond to the regime of Lifshitz
localization, in the tail of the density of states and controlled by a
rare-events scenario. Adapting Lifshitz argument~\cite{Lifshitz1965,
  Luck1992}, a region of length $\ell$ with many $\tilde{\beta}_j$
close to $\beta_{\text{min}}$ -- i.e. provided
$\modul{\tilde{\beta}_j-\beta_{\text{min}}}\leq
C(E_{\ell}-E_{\text{min}})$ for all $j$ in the region, with $C$ a
constant -- has its lowest energy of the order of $E_{\ell} \simeq
E_{\text{min}} + \beta_{\text{min}}\pi^2/2\ell^2$, assuming
$\beta_{\text{min}}>0$ and writing $E_{\text{min}} =
\beta_{\text{min}}/4$. Since the probability of creating such region
scales as $P_{\ell} \propto
[C(E_{\ell}-E_{\text{min}})/2\Delta]^{\ell}$ for a uniform
distribution, the low-energy behavior of the integrated density of
states is
\begin{equation}
N(\energy) \propto \exp\Bigg\{-\pi\sqrt{\frac{\beta_{\text{min}}/2}{\energy-\energy_{\text{min}}}}\ln\Bigg(\frac{2\Delta/C}{\energy-\energy_{\text{min}}}\Bigg)\Bigg\}\;.
\label{eq:lifshitz}
\end{equation}
This behavior is in very good agreement with the numerics on the
effective model~\cite{supplementary}. Regarding finite-size effects on
the spinon energy $\energy_L$ in a chain of length $L$, the
probability to have the minimum energy must be such that $P_{\ell}
\sim 1/L$. This yields $N(\energy_L) \sim 1/L$ in
Eq.~\eqref{eq:lifshitz}, again in good agreement with numerical
results~\cite{supplementary}. Asymptotically, we thus expect
finite-size corrections of the form $\energy_L \simeq
\energy_{\text{min}} + K \beta_{\text{min}}(\ln(\ln L)/\ln L)^2$ with
$K$ a constant.

In order to validate this effective model, we compare it to accurate
DMRG calculations of the magnetization profile in a chain with total
spin $S^z=1/2$ with the effective model predictions where $m_i \equiv
\moy{S^z_i}$ is deduced from the $\psi_j$~\cite{supplementary}. As
judged by the results of Fig.~\ref{fig:localization}(b), the effective
model provides quantitative predictions of the magnetization profile.
>From the local dimerization pattern $d_i =
\moy{\mathbf{S}_i\cdot\mathbf{S}_{i+1}}$ of
Fig.~\ref{fig:localization}(c), the localized spinon clearly separates
two different MG domains. One can extract the actual
$\xi_{\text{RMG}}$ from DMRG profiles. On
Fig.~\ref{fig:localization}(d), strong deviations between DMRG
calculations and the infinite size result \eqref{eq:xi-weak-coupling}
are observed. This difference actually originates from finite-size
effects : one has to take $\xi_{\text{RMG}} = 1/\gamma(\energy_L)$
on a finite system, which yields strong deviations, even for large
sizes.

\begin{figure}[t]
\centering
\includegraphics[width=\columnwidth,clip]{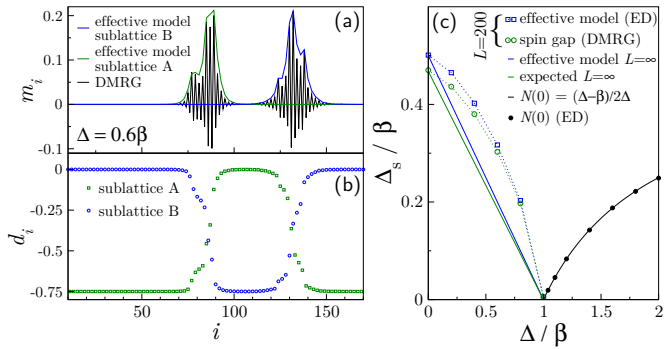}
\caption{(Color online) \emph{At the RMG point} -- typical
  magnetization (a) and dimerization (b) profiles in the lowest
  triplet excited state for $\Delta<\beta$. (c) evolution of the
  minimum spin gap (see text for details), and spinon density $N(0)$,
  v.s. the disorder strength $\Delta$.}
\label{fig:gap}
\end{figure}

\emph{Spin gap and transition to a paramagnetic state} -- From these
results on single spinon excitations, we can infer the behavior of the
spin gap at the RMG point with increasing disorder. The MG states
remain degenerate ground-states at small enough disorder and the
lowest triplet excitation above them is to create two localized
spinons in each sublattice (see Fig.~\ref{fig:gap}(a-b)) with both the
minimal energy $E_{\text{min}}$. As the two spinons are localized on
very large clusters, the singlet and triplet gaps must be degenerate
in the thermodynamical limit~\cite{footnote}. Consequently, the
effective model prediction for the spin gap is $\Delta_S^{\text{eff}}
= \beta_{\text{min}}/2 = (\beta-\Delta)/2$. In order to give a typical
finite-size behavior, we show on Fig.~\ref{fig:gap}(c) the minimum
DMRG triplet gap found over an hundred of samples of chains of size
$L=200$, compared with the effective model result on the same
samples. Two main features come out. First, the difference between the
two finite-size curves can be attributed to the variational error,
already making the non-disordered spin gap $\Delta_S^0$ smaller than
$\Delta_S^{\text{eff}}$. As Lifshitz states correspond to large
`clean' boxes with couplings $\beta_{\text{min}}$, the same correction
should apply to them, improving the prediction to $\Delta_S =
\Delta_S^0(1-\Delta/\beta)$ (label `expected $L=\infty$' on
Fig.~\ref{fig:gap}(c)). Second, the curved nature of the typical spin
gap for $L=200$ and randomly chosen samples is due to the fact that
the minimum gap discussed above is obtained for extremely rare events
and can be viewed as a sort of finite-size effect since $\Delta_S$ is
not self-averaging.

Increasing further the disorder strength, the spin gap vanishes for
$\beta=\Delta$ for which $E_{\text{min}}$ becomes negative
($E_{\text{min}}=9\beta_{\text{min}}/4$ for $\beta_{\text{min}}<0$) so
that states with two or more spinons get energetically favored. The MG
states are still eigenstates but no longer ground-states. The
resulting picture shortly after the critical point is a paramagnetic
phase of localized spinons, as we can neglect tiny residual magnetic
couplings between spinons for a small enough density. This spinon
density is nothing but the density of negative energy spinon states
$N(0)$. Therefore, the magnetic susceptibility must change from zero
to $\chi \sim N(0)$ across the transition as depicted in
Fig.~\ref{fig:gap}(c). DMRG calculations do confirm this picture,
showing that the ground and first excited states are nearly degenerate
states with localized spinons. Within the effective model picture, the
quantum phase transition from the gapped to the paramagnetic phase is
located at $\beta_{\text{min}}=0$ and its order depends on the
disorder distribution. A continuous transition occurs for a continuous
disorder distribution while binary disorder yields a first order
transition. In order to determine $N(0)$ for the effective model, one
must realize that the Lifshitz argument cannot be used for spinon
energies close to zero~\cite{supplementary}. Yet, we argue that the
number of negative energy states is actually given by the number of
negative $\tilde{\beta}_j$ which, for the box distribution, gives
$N(0)=(\Delta-\beta)/(2\Delta)$. The associated critical exponent of
the susceptibility is thus one. This argument is checked numerically
on the effective model in Fig.~\ref{fig:gap}(c). Checking this law
using DMRG is particularly difficult as the spinon density gets very
small close to the critical point. Lastly, we point out that this low
spinon density picture fails at larger disorder (large spinon density)
for two related reasons: magnetic couplings between spinons become
non-negligible and the neglecting states with non-local dimers is
questionable. It is likely that the spinon phase then becomes
partially polarized and connected to the large-spin phase that exists
away from the RMG point and that we now discuss.

\emph{Random confinement localization} -- Moving away from the RMG
point \eqref{eq:MGcondition} is progressively done by uncorrelating
the $\alpha_i$ and $\beta_i$ through the introduction of random
variables $\delta_i$, uncorrelated to the $\beta_i$, and such that
\begin{equation}
\alpha_i = (1-\lambda)(\beta_i+\beta_{i+1}) + \lambda\delta_i\;,
\label{eq:uncorrelated}
\end{equation}
where $\overline{\delta_i} = 2\beta$ and $\lambda \in [0,1]$ is a
tuning parameter. The correlations
$\overline{\alpha_i\beta_i}-\overline{\alpha_i}\overline{\beta_i} =
(1-\lambda) \sigma_{\beta}^2$ ($\sigma^2$ denoting a variance) show
that the $\alpha$s and $\beta$s get uncorrelated for $\lambda=1$.  As
$\sigma_{\alpha}^2 = 2(1-\lambda)^2\sigma_{\beta}^2 + \lambda^2
\sigma_{\delta}^2$, we further impose $\sigma_{\alpha}^2 =
2\sigma_{\beta}^2$ to study the effect of the correlations only by
keeping the disorder strength as for the RMG point. Lastly, using this
decoupling, the Hamiltonian splits into two parts $\Ham =
\Ham_{\text{RMG}} + \lambda \Ham_{\text{dim}}$, where $\Ham_{\text{dim}}$
is a random dimerization term:
\begin{equation}
\Ham_{\text{dim}} = \sum_i \eta_i \mathbf{S}_i\cdot\mathbf{S}_{i+1}\;,
\label{eq:Hdim}
\end{equation}
with $\eta_i = \delta_i - \beta_i -\beta_{i+1}$ random couplings of
mean-value $\overline{\eta_i} = 0$. The effect of $\Ham_{\text{dim}}$
is to localize spinons via a random confinement mechanism that can be
understood within the effective model approach. The associated
effective Hamiltonian $\widetilde{\Ham}_{\text{dim}}$ in the single
spinon basis takes a rather complicated form~\cite{supplementary},
with a dense-matrix representation. Still, its main effect is to
induce an effective random chemical potential $\mu_i$ for the spinon which
reads
\begin{equation}
\mu_i = \elem{i}{\Ham_{\text{dim}}}{i}=-\frac{3}{4}\bigg(\sum_{n=0}^{i-1}\eta_{2n+1}+\sum_{n=i+1}^{(L-1)/2}\eta_{2n}\bigg)\;.
\label{eq:Edim}
\end{equation}
It is the sum of two independent random walks (the $\eta$s on each
sublattice) constrained to have a total fixed length. Such a potential
typically has a minimum in the bulk of large chains and create a well
which localizes the spinon. As for explicitly dimerized
chains~\cite{DimerizedChain, Uhrig1999}, moving away from the minimum
of the potential is locally analogous to a linear confinement. DMRG
calculations are nicely fitted by diagonalizing the full effective
hamiltonian $\widetilde{\Ham}$ matrix~\cite{supplementary}. The
location of the spinon is found to be governed by the $\mu_i$.
If one now takes the full Hamiltonian, finite-size systems display a
crossover between Anderson localization at very small $\lambda$ and
random confinement localization which dominates up to $\lambda=1$. It
is also clear that the random confinement can stabilize many spinons
states leading to the formation of many domains.

\emph{MG domain formation} -- At small $\lambda$, $\Ham_{\text{dim}}$
acts as a perturbation which locally lifts the degeneracy between the
two MG states, as suggested qualitatively in
Ref.~\onlinecite{Yang1996}.  Clearly, MG states are no longer
eigenstates since high order terms in perturbation theory put weights
on long dimer states $\ket{[2i-1,2j]} = \ket{\MGstateLongEx}$ with a dimer on bond
$(2i-1,2j)$. More than dressing the MG state, the random dimerization
actually destroys the spin gap as soon as $\lambda\neq 0$ and
$\sigma_{\beta} \neq 0$. Indeed, the energy difference $\delta E_{ij}
= \elem{[2i-1,2j]}{\Ham}{[2i-1,2j]}-E_{\text{MG}}$ between $\ket{[2i-1,2j]}$ and
the corresponding MG state reads~\cite{supplementary}
\begin{equation}
\delta E_{ij} =\frac{3}{4}\big(\beta_{2i-1}+\beta_{2j}\big)-\lambda\frac{3}{4}\sum_{n=2i-1}^{2j-1}(-1)^{n}\eta_{n}.
\label{eq:potenergy}
\end{equation}
The first term which averages to $\frac 3 2\beta$ stems from the cost of
creating two domain walls (spinons). The second term corresponds to an
effective long range interaction between the two spinons and arises
from $\Ham_{\text{dim}}$. It averages to zero but rare events can
definitely bring this state to a lower energy than the MG state: in an
infinitely large system, it is always possible to find a region with
$\eta$s that do not compensate and such that the random interaction
scales as the region size to make $\delta E_{ij} < 0$. At low
disorder, regions must be large and the spinons far away so that their
remaining magnetic coupling vanishes. The spin gap is then immediately
broken.
 
\begin{figure}[t]
\centering
\includegraphics[width=\columnwidth,clip]{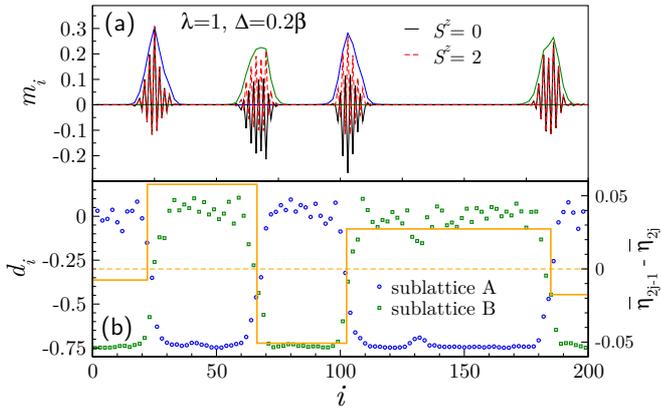}
\caption{(Color online) (a) Magnetization profile of a MG chain with
  uncorrelated couplings displaying localized spinons. Lines are the
  effective model predictions on each sublattices. (b) Dimerization
  pattern and corresponding average of the random $\eta$-dimerization
  term of Eq.~\eqref{eq:potenergy} over MG domains (orange line).}
\label{fig:finitelambda}
\end{figure}

In order to support this picture, we perform DMRG calculations for
$\lambda\neq 0$ and do observe that, on a finite-system, the spin gap
is strongly reduced by increasing $\lambda$, or by increasing $\Delta$
at fixed $\lambda$. We exhibit in Fig.~\ref{fig:finitelambda}(a) a
sample with $\lambda=1$ where four spinons are present in the
ground-state and for which the spin gap is zero within numerical
accuracy. To complete the description, the magnetization profile along
each domain wall is pretty well reproduced by using the local
effective Hamiltonian for a single spinon (Born-like approximation).
Finally, the averaged of $\Ham_{\text{dim}}$ in each domain, plotted
in Fig.~\ref{fig:finitelambda}(b), supports the pining mechanism of
the MG domains. The scaling of the spinon density with disorder is a
challenging issue as it requires the minimization of the energy of
several correlated domains which number is not fixed~\cite{footnote2}.

At low spinon densities, the ground-state is a network of localized
spinons which interact via their spin degrees of freedom. The
effective magnetic couplings should range from almost zero to finite
values if two spinons happen to be close. There is no constraint on
their signs: both ferromagnetic and anti-ferromagnetic couplings
exist, making the phase partially polarized. Interestingly this weak
disorder picture is physically connected with the strong disorder
picture of RSRG.

\emph{Comparison with RSRG} -- It is interesting to compare these
results with the RSRG method best suited to the strong disorder
limit. From the RSRG equations given in the supplementary material, we
notice that the degeneracy of the MG domains at the RMG point
translates into an instability of the RSRG decimation. As soon as
$\lambda\neq 0$, the gap distribution in the system converges toward
an invariant power-law distribution with a non-universal exponent,
caracteristic of Griffith phase similar to previous
results~\cite{Hoyos2004}. Indeed, due to frustration, the RSRG
equations generate a few effective ferromagnetic
couplings~\cite{supplementary}, building up a large-spin
phase~\cite{Westerberg1997}. This supports a continuous phase from
weak to strong disorder. Moreover, we stress that the non-crossing
dimer basis is deeply related to the RSRG picture that targets the
most probable dimer configuration from the coupling distribution.

\emph{Conclusion} -- This work provides quantitative results on the
interplay between frustration and disorder in random MG chains. We
identify two mechanisms at play for single spinon: an unsual Anderson
localization mechanism, and a random confinement mechanism. The
immediate destruction of the spin gap upon putting disorder is to be
contrasted with its robustness for the explicitly dimerized or spin-1
chains which have a non-degenerate ground-state~\cite{Hyman1996}. The
presence of degenerate MG states makes the system very sensitive to
disorder. We expect the same phenomenology to play a role in other
random VBS, which could be witnessed using numerical methods working
in the spinon basis~\cite{Tang2011}. Lastly, these mechanisms
demonstrate how random couplings generate ``free spins'' in VBS,
without vacancies or adatoms, and are thus experimentally relevant.

We thank N. Laflorencie, J.-M. Luck, C. Monthus, C. Sire and C. Texier
for insightful discussions. We acknowledge support from grant
ANR-2011-BS04-012-01 QuDec.

\cleardoublepage
\onecolumngrid
\appendix

\begin{center}
{\Large\textbf{Supplementary material for: Localization of spinons in random Majumdar-Ghosh chains}}
\end{center}

\section{Random variables features}

For clarity, we list below the random variables that appear in the
study, as well as their mean-value, variance and correlations with the
$\beta_i$ variable:
\begin{center}
\begin{tabular}{|c|c|c|c|}
\hline
variable & mean & variance $\sigma^2$ & correlations \\
\hline\hline
$\beta_i$  & $\beta$ &  $\sigma_\beta^2$ ($ = \Delta^2/3 $ for the box distribution) &   \\
$\alpha_i = (1-\lambda)(\beta_i+\beta_{i+1}) + \lambda\delta_i$ & $\alpha=2\beta$ &  $\sigma_\alpha^2 = 2 \sigma_\beta^2$\quad (by choice)  &  $ \overline{\alpha_i\beta_i} -\alpha\beta = (1-\lambda)\sigma_\beta^2$  \\
$\delta_i$ & $\delta = 2\beta$ &  $\sigma_\delta^2 = 2\frac{2-\lambda}{\lambda}\sigma_\beta^2$\quad (by choice of $\sigma_\alpha^2$) &  $\overline{\delta_i\beta_i} -\delta\beta =0$ \\
$\eta_i = \delta_i-\beta_i-\beta_{i+1}$   & $0$      & $\sigma_\eta^2 = \sigma_\delta^2+2\sigma_\beta^2 = \frac{4}{\lambda}\sigma_\beta^2$ & $\overline{\eta_i\beta_i} = -\sigma_\beta^2$ \\
\hline\hline
\end{tabular}
\end{center}

\section{Non-crossing dimer basis}
\label{app:dimerbasis}

\subsection{$S_\text{tot}=0$ sector and MG states}

We gather some useful results on the non-crossing dimer basis used for
variational calculations. Non-crossing dimer states form a
non-orthogonal basis of the subspace with total spin
$S_\text{tot}=0$. In the MG physics, the states with dominant weights
are rather simple as they are essentially states with nearest-neighbor
dimers, with possibly slightly longer dimers locally. We also recall
how the different terms of the Hamiltonian act on a MG state
$\ket{\text{MG}} = \ket{\MGstate}$ with dimers $\ket{\dimer} =
\frac{1}{\sqrt{2}}[\ket{\ups\downs}-\ket{\downs\ups}]$. Applying a
nearest neighbour term on a dimer, one simply recovers the MG state
with eigenvalue $-3/4$. The same term applied between two dimers
gives:
\begin{align}
\mathbf{S}_i\cdot\mathbf{S}_{i+1}\ket{\MGstate} = \frac{1}{4}\ket{\MGstate} + \frac{1}{2}\ket{\MGstateEx}
\end{align}
Applying a next-nearest neighbour term, one gets :
\begin{align}
\mathbf{S}_i\cdot\mathbf{S}_{i+2}\ket{\MGstate} = \frac{1}{4}\ket{\MGstate} - \frac{1}{2}\ket{\MGstateExCross},
\label{eq:MG}
\end{align}
which can be rewritten in the non-crossing dimer basis, using
\begin{align}
\ket{\MGstateExCross}= \ket{\MGstate}  +\ket{\MGstateEx}.
\end{align}
A useful overlap that will often appear in calculation is:
\begin{equation}
\langle{\MGstate}\ket{\MGstateEx}=-1/2\;.
\label{eq:overlap}
\end{equation}

\subsubsection{Deriving the Random Majumdar-Ghosh condition}

Applying $\Ham$ on the state $\ket{\text{MG}}$ which starts on even
sites $2j$ gives
\begin{align*}
\mathcal{H}\ket{\text{MG}}&= \frac 1 4 \sum_j \left( -3 \alpha_{2j} + \alpha_{2j+1} - \beta_{2j} - \beta_{2j+1}\right) \ket{\text{MG}} 
+ \frac 1 2 \sum_j \left(\alpha_{2j-1} - \beta_{2j} - \beta_{2j-1}\right) \ket{[2j-2,2j+1]}\;,
\end{align*}
where we write $\ket{[2j-2,2j+1]} = \ket{\MGstateEx}$, the state with a
dimer on bond $(2j-2,2j+1)$. A similar expression is obtained for the
other MG state. The RMG point is obtained by cancelling the second
term.

\subsubsection{Some features of the Majumdar-Ghosh state}
\label{app:MGstate}

We discuss some remarkable features of the MG state $\ket{\text{MG}}$
in the presence of disorder. From \eqref{eq:MG} and \eqref{eq:overlap}
and using and that $\sum_j (\cdots) \rightarrow
\frac{L}{2}\overline{(\cdots)}$ for large enough system
\begin{equation}
\bra{\text{MG}}\mathcal{H}\ket{\text{MG}} = E_{\text{MG}} = -\frac{3}{4}\sum_{j=1}^{L/2} \alpha_{2j-1} \rightarrow -\frac 3 4 L \beta \;.
\label{eq:EMG}
\end{equation}
Notice that the above energy is exact even when \eqref{eq:MGcondition}
is not satisfied, but that the simplification from averaging only
comes with the thermodynamic limit. In particular, it is remarkable
that the energy remains independent of the disorder strength $\Delta$.
The fact that the MG state is not an eigenstate in general (when
\eqref{eq:MGcondition} is not satisfied) can be captured by
calculating the energy dispersion of the state:
\begin{eqnarray}
\sigma_{\text{MG}}^2 \equiv \bra{\text{MG}}\mathcal{H}^2\ket{\text{MG}} - E^2_{\text{MG}}
                    \rightarrow \frac{3L}{16}\left[\sigma_{\alpha}^2 + 2\sigma_{\beta}^2 + 2\alpha^2 + 8 \beta^2 - 4(\overline{\alpha\beta}+\alpha\beta)\right] = \lambda \frac 3 4 L \sigma_{\beta}^2
\label{eq:sigmaMG}
\end{eqnarray}
Clearly, we do have $\sigma_{\text{MG}} = 0$ when
\eqref{eq:MGcondition} is fulfilled, as expected for an eigenstate.

\subsection{$S_\text{tot}=1/2$ sector and single spinon states}

The single spinon dynamics is obtained using the restriction of the
Hamiltonian on the subspace of states $\ket{i}=\ket{\spinon}$ with
spinon at position $2i+1$. We assume an infinite length chain, or a
finite odd size chain with open boundary conditions, so that the
spinon only moves on one sublattice. Periodic boundary conditions
would only allow the spinon to change sublattice at the edges of the
chain. Therefore it would just lead to a doubling of the effective
size of the chain for the spinon. This method is variational as the
family of states $\ket{i}$ does not form a complete family of subspace
$\{S_\text{tot}=1/2, S^z_\text{tot}=1/2\}$. Moreover this family is
free but is not orthogonal. The overlap between two states is
\begin{equation}
\braket{i}{j}=\left(-\frac{1}{2}\right)^{|i-j|}.
\end{equation}
Magnetization profiles $m_i =
\dfrac{\elem{\psi}{S^z_i}{\psi}}{\braket{\psi}{\psi}}$ can easily be
computed~\cite{Uhrig1999} in this basis using :
\begin{align}
\braket{\psi}{\psi} & =\sum_{j,k}\psi_k^*\psi_j \braket{k}{j},\\
\elem{\psi}{S^z_{2i+1}}{\psi} & =\frac{1}{2}\sum_{k\leq i\leq j} \psi_k^*\psi_j \braket{k}{j}+\text{c.c.}, \\
\elem{\psi}{S^z_{2i}}{\psi} & =-\frac{1}{2}\sum_{k< i\leq j} \psi_k^*\psi_j \braket{k}{j}+\text{c.c.},
\end{align}
with $\ket{\psi} = \sum_i \psi_i \ket{i}$. If $|\psi_i|$ varies slowly
compared to $(1/2)^i$, that is if the localization length is large
($\xi_\text{spinon\ }\gg 1/\ln2$), the above sums can be approximated:
\begin{align}
\braket{\psi}{\psi} & \simeq 3\sum_i|\psi_i|^2, \\
\elem{\psi}{S^z_{2i+1}}{\psi} & \simeq \frac{7}{2}|\psi_i|^2,\\
\elem{\psi}{S^z_{2i}}{\psi} & \simeq -\left(|\psi_i|^2+|\psi_{i-1}|^2\right).
\end{align}
In practice, these expressions work really well even for short
localization lengths, and we did not need to compute the exact
magnetization profiles. As the effective Hamiltonians are reals, all
$\psi_i$ are actually real numbers in our case.

\section{Effective Hamiltonian for spinons}
\label{app:effectivehamiltonian}

\subsection{Projection on the variational subspace}

We call $P$ the orthogonal projector on the subspace generated by
states $\ket{i}$. As an orthogonal projector, $P$ is self-adjoint. The
effective Hamiltonian for a single spinon $\widetilde{\Ham}$ is the
restriction of $\Ham$ on this subspace :
\begin{equation}
\widetilde{\Ham}=P\Ham P
\end{equation}
We want to diagonalize $\widetilde{\Ham}$ that is to find the energies
$E$ and the eigenstates $\ket{\psi} = \sum_i \psi_i \ket{i}$ so that
\begin{equation}                                                                                                                                            
\widetilde{\Ham}\ket{\psi}=E\ket{\psi}.
\end{equation}
Of course, as $\Ham$ is self-adjoint, $\widetilde{\Ham}$ is also
self-adjoint and the variational energies are real. If we write
$\ket{\psi}$ as a variational wavefunction $\ket{\psi}=\sum_i \psi_i
\ket{i}$, diagonalizing $\mymat{\widetilde{\Ham}}_{ij}$ the matrix of
$\widetilde{\Ham}$ in the basis of states $\ket{i}$ is equivalent to
solve the generalized eigenvalue problem :
\begin{equation}
\sum_i\elem{j}{\Ham}{i}\psi_i = E\sum_i\braket{j}{i}\psi_i.
\end{equation}
We insist on the fact that the matrices $\elem{j}{\Ham}{i}$ and
$\mymat{\widetilde{\Ham}}_{ij}$ are different because the basis of
states $\ket{i}$ is not orthogonal. Indeed, it is useful in this
context to introduce the matrix of overlaps $\mathcal{O}$ which
elements are $\mymat{\mathcal{O}}_{ij}=\braket{i}{j}$.  The search for
eigenvalues in the generalized eigenvalue problem takes the form
$\det(\Ham - E \mathcal{O})=0$ which is equivalent to
$\det(\mathcal{O}^{-1}\Ham - E \mathbb{I})=0$. In addition, the
projector on the subspace $\{\ket{i}\}$ is the inverse of the overlap
matrix, ie. $P = \sum_{ij} \mymat{\mathcal{O}^{-1}}_{ij}
\ket{i}\bra{j}$.  Then, $\mymat{\widetilde{\Ham}}_{ij}$ is deduced
from $\elem{i}{\Ham}{j}$ by
\begin{equation}
\mymat{\widetilde{\Ham}}_{ij} = \sum_k \mymat{\mathcal{O}^{-1}}_{ik}\elem{k}{\Ham}{j}.
\end{equation}
In the case of a chain, the inverse of $\mathcal{O}$ has a simple
tridiagonal form:
\begin{equation} \mymat{\mathcal{O}^{-1}}_{ij} = \frac 1 3
\begin{pmatrix}
 4 & 2 &   &   &   &    \\
 2 & 5 & 2 &   &   &    \\
   & 2 & 5 & 2 &   &    \\
   &   & \ddots & \ddots & \ddots &    \\
   &   &        &   2    &   5   & 2   \\
   &   &        &       &    2 &  4 \\
\end{pmatrix}\;,
\end{equation}
which allows one to treat the problem analytically.

\subsection{Effective Hamiltonian}

We now detail how the terms of the Hamiltonian act on a single spinon
state $\ket{i}$. Applying $\mathbf{S}_j\cdot\mathbf{S}_{j+1}$ on a
spinon at position $j$, one gets
\begin{align}
\mathbf{S}_j\cdot\mathbf{S}_{j+1}\ket{\spinon} = \frac{1}{4}\ket{\spinon} + \frac{1}{2}\ket{\spinonHopeRight}
\end{align}
Using the following relation
\begin{align}
\ket{\spinonExcitedRoofRight}=\ket{\spinon}+\ket{\spinonHopeRight}, 
\end{align}
one can deduce the application of $\mathbf{S}_j\cdot\mathbf{S}_{j+2}$
in the variational basis :
\begin{align}
\mathbf{S}_j\cdot\mathbf{S}_{j+2}\ket{\spinon} = -\frac{1}{4}\ket{\spinon} - \frac{1}{2}\ket{\spinonHopeRight}
\end{align}
Applying $\mathbf{S}_{j-1}\cdot\mathbf{S}_{j+1}$ on a spinon at
position $j$, one gets :
\begin{align}
\mathbf{S}_{j-1}\cdot\mathbf{S}_{j+1}\ket{\spinon} = \frac{1}{4}\ket{\spinon} + \frac{1}{2}\ket{\spinonExcited}
\end{align}
One can notice that this state is orthogonal to the variational
subspace. As a result it simply disappears within the variational
approach.

\subsection{Calculation of $\widetilde{\Ham}_{\text{RMG}}$}

Finally, the RMG Hamiltonian applied on a spinon at position $i$ and
projected on the variational subspace, gives :
\begin{align}
\widetilde{\Ham}_{\text{RMG}}\ket{i} = \frac{1}{4}\left(-3\sum_j \beta_j+5\beta_{2i+1}\right)\ket{i}
                  + \frac{1}{2}\beta_{2i+1}\ket{i+1}
                  + \frac{1}{2}\beta_{2i+1}\ket{i-1}                 
\label{eq:Hrmg}
\end{align}

\subsubsection{Similarity to a symmetric matrix}

The effective model of Eq.~\eqref{eq:spinon-Ham} is in a non-hermitian
form due to the non-orthogonal nature of the dimer basis. The
associated matrix in the variational basis simply reads (writing
$s=5/2$):
\begin{equation} \mymat{\widetilde{\Ham}_{\text{RMG}}}_{ij} = \frac 1 2
\begin{pmatrix}
s\beta_1 &  \beta_1 &         &          &          \\
\beta_2  & s\beta_2 & \beta_2 &          &         \\
         & \beta_3  & s\beta_3& \beta_3  &          \\
         &          & \beta_4 & s\beta_4 & \ddots \\
         &          &         & \ddots   & \ddots \\
\end{pmatrix}\;.
\end{equation}
\emph{When all the $\beta$s are positive numbers}, using the
similarity transform diagonal matrix $D =
\text{diag}(\sqrt{\beta_1},\sqrt{\beta_2},\ldots)$ puts
$\widetilde{\Ham}_{\text{RMG}}$ into the following tridiagonal
symmetric form
\begin{equation}
\mymat{D^{-1}\widetilde{\Ham}_{\text{RMG}}D}_{ij} = \frac 1 2
\begin{pmatrix}
s\beta_1 &  \sqrt{\beta_1\beta_2} &         &          &          \\
\sqrt{\beta_1\beta_2} & s\beta_2 & \sqrt{\beta_2\beta_3} &          &         \\
         & \sqrt{\beta_2\beta_3}  & s\beta_3& \sqrt{\beta_3\beta_4}  &          \\
         &          & \sqrt{\beta_3\beta_4} & s\beta_4 & \ddots \\
         &          &         & \ddots & \ddots \\
\end{pmatrix}\;.
\end{equation}
clearly showing that all eigenvalues are real, as for the
eigenvectors.

\emph{When some of the $\beta$s are negative}, applying the same
transform then leads to a complex symmetric matrix, but not to an
hermitian one. A non-diagonal similarity transform is then required
which would make in general the hermitian matrix dense. Looking at the
$2\times2$ and $3\times3$ cases shows that it becomes pretty difficult
to construct a similarity transform matrix. Still, the discussion on
the projection method for obtaining the effective hamiltonian above
ensures us that the spectrum is real, in agreement with numerical
results.

\subsubsection{Weak-coupling results for $\widetilde{\Ham}_{\text{RMG}}$}

We gather the weak-coupling results on the Lyapunov exponent of the
effective model. The energy is parametrized through the variables $t$
or $k$:
\begin{itemize}
\item for $E<\frac{\beta}{4}$ and $E=\beta(\frac{5}{4}-\cosh t)$:\quad $\displaystyle \gamma=t-\frac{1}{6}\left(\frac{(\frac{5}{4}-\cosh t)\Delta}{\sinh(t)\beta}\right)^2$
\item for $\frac{\beta}{4}<E<\frac{9\beta}{4}$ and $E=\beta(\frac{5}{4}+\cos k)$:\quad $\displaystyle \gamma=\frac{1}{6}\left(\frac{(\frac{5}{4}+\cos k)\Delta}{\sin(k)\beta}\right)^2$
\item for $E>\frac{9\beta}{4}$ and $E=\beta(\frac{5}{4}+\cosh t)$: \quad $\displaystyle \gamma=t-\frac{1}{6}\left(\frac{(\frac{5}{4}+\cosh t)\Delta}{\sinh(t)\beta}\right)^2$
\end{itemize}

\subsubsection{Numerics on the Lifshitz tail}

\begin{figure}[t]
\centering
\includegraphics[width=0.65\columnwidth,clip]{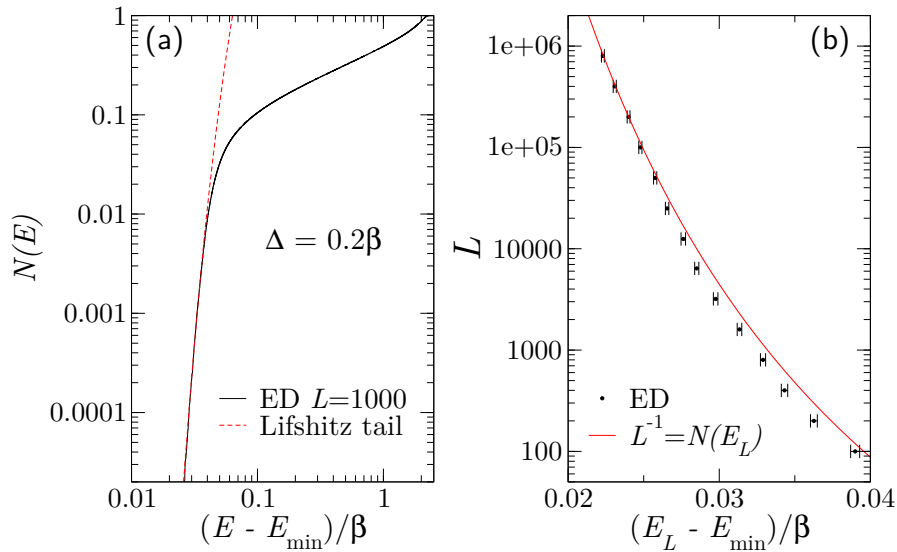}
\caption{(Color online) (a) Integrated density of states at
  low-energies compared with Lifshitz argument od
  Eq.~\eqref{eq:lifshitz}. (b) Finite-size effects on the spinon
  energy.}
\label{fig:NE}
\end{figure}

We give in Fig.~\ref{fig:NE}(a-b) the comparison between numerical
calculations on the effective model and finite-size corrections
obtained from Lifshitz argument.

\subsubsection{A wrong argument for the susceptibility exponent at the random MG point}

A naive argument for $\beta_{\text{min}}<0$ is the following : the
susceptibility $\chi$ of the paramagnetic phase should correspond to
independently filling spinons in single-spinon \emph{Lifshitz states}
up to zero energy. One is tempted to use a Lifshitz formula for
$N(0)$, which is similar to \eqref{eq:lifshitz} with the changes
$E_{\ell} \simeq E_{\text{min}} - \beta_{\text{min}}\pi^2/2\ell^2$ and
$E_{\text{min}} = 9\beta_{\text{min}}/4$. Using $\beta_{\text{min}} =
\Delta_c-\Delta$, with $\Delta_c=\beta$ the critical disorder
strength, one gets for the susceptibility exponent $\phi =
\pi\sqrt{2}/3 \simeq 1.481$ for the uniform distribution. Actually,
such a prediction is wrong for the reason that the Lifshitz formula
does not work close to the $E=0$ while it does work close to
$E_{\text{min}}$. Indeed, when $\beta_{\text{min}}\simeq0$, the energy
$E_{\ell}$ of a state in a cluster of size $\ell$ no longer depends on
$\ell$.  This behavior is true whatever the smallness of the disorder
strength. Instead, we observe that $N(E)$ is linear close to $E=0$ and
that $N(0)$ is rather linear with $\Delta-\Delta_c$. As discussed in
the main text, the correct $N(0)$ is obtained by coupling the number
of negative $\beta$s which leads to $\phi=1$ for the uniform
distribution.

\subsection{Calculation of $\widetilde{\Ham}_{\text{dim}}$}

If we now want to do the same for the dimerization term of the
Hamiltonian:
\begin{align}                                                                                                                                               \Ham_{\text{dim}}=\sum_i \eta_i \mathbf{S}_i\cdot\mathbf{S}_{i+1},
\end{align}
we have to consider states $\ket{[2n-1,2n+2],2i+1}$ ($n<i$) with a
spinon at site $2i+1$ and a dimer on bond $(2n-1,2n+2)$, and states
$\ket{2i+1,[2n,2n+3]}$ ($n>i$) with a spinon at site $2i+1$ and a
dimer on bond $(2n,2n+3)$.  The overlaps of these states with states
$\ket{j}$ are :
\begin{align}
\braket{j}{[2n-1,2n+2],2i+1}  &=-\frac{1}{2}\braket{j}{i}(1+3\theta(n-j-1))\\
\braket{j}{2i+1,[2n,2n+3]} &=-\frac{1}{2}\braket{j}{i}(1+3\theta(j-n-1))
\end{align}
where $\theta$ is the Heaviside step function with the choice
$\theta(0)=1$. These states can be projected on the variational
subspace.
\begin{align}
P\ket{[2n-1,2n+2],2i+1} &=-\frac{1}{2}\ket{i}-\left(-\frac{1}{2}\right)^{i-n+1}\ket{n}+\left(-\frac{1}{2}\right)^{i-n}\ket{n-1} \\
P\ket{2i+1,[2n,2n+3]} &=-\frac{1}{2}\ket{i}-\left(-\frac{1}{2}\right)^{n-i+1}\ket{n}+\left(-\frac{1}{2}\right)^{n-i}\ket{n+1}
\end{align}
Using these results, one can obtain the expression of
$\widetilde{\Ham}_{\text{dim}}$ in the variationnal basis :
\begin{align}
\widetilde{\Ham}_{\text{dim}}\ket{i}=
&\left[-\frac{3}{4}\left(\sum_{n=0}^{i-1}\eta_{2n+1}+\sum_{n=i+1}^{\frac{L-1}{2}}\eta_{2n}\right)+\frac{1}{4}\left(\eta_{2i}+\eta_{2i+1}\right)\right]\ket{i} \nonumber \\
&+\frac{1}{2}\eta_{2i}\ket{i-1}+\frac{1}{2}\eta_{2i+1}\ket{i+1}\nonumber\\
&+\sum_{n=1}^{i-1}\left(-\frac{1}{2}\right)^{i-n}\eta_{2n}\left(\frac{1}{4}\ket{n}+\frac{1}{2}\ket{n-1}\right)\nonumber \\
&+\sum_{n=i+1}^{\frac{L-3}{2}}\left(-\frac{1}{2}\right)^{n-i}\eta_{2n}\left(\frac{1}{4}\ket{n}+\frac{1}{2}\ket{n+1}\right)
\end{align}
The energie due to the random dimerization $\eta$ of a spinon
localized at site $2i+1$ is :
\begin{align}
\elem{i}{\Ham_{\text{dim}}}{i}=-\frac{3}{4}\left(\sum_{n=0}^{i-1}\eta_{2n+1}+\sum_{n=i+1}^{\frac{L-1}{2}}\eta_{2n}\right)
\end{align}

\subsection{Numerical checks of one spinon in an open chain}

\begin{figure}[b]
\centering
\includegraphics[width=0.7\columnwidth,clip]{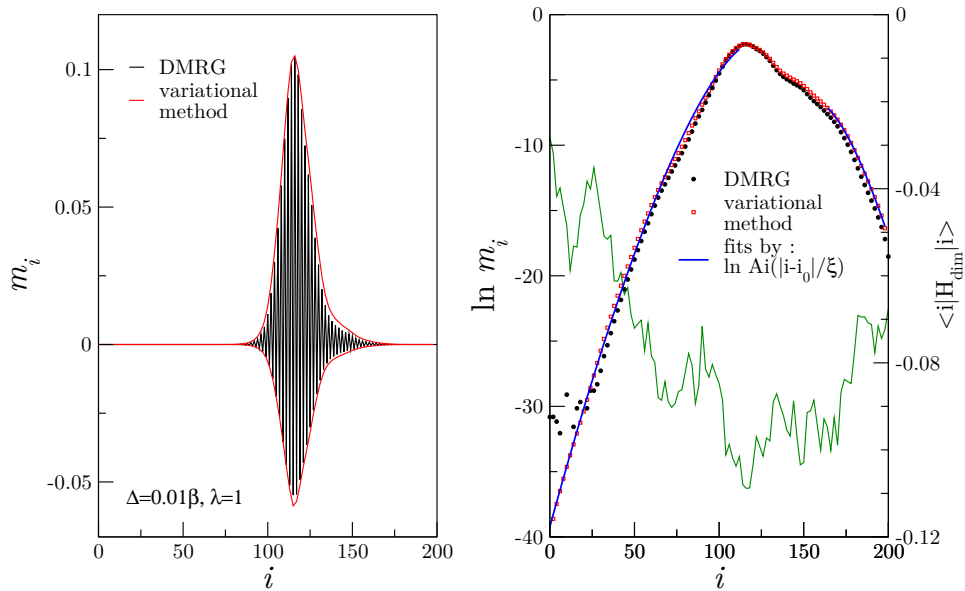}
\caption{(Color online) (a) Magnetization profile of a single spinon
  wave-function away from the RMG point. (b) Log plot showing fits by
  Airy functions and the effective local potential
  $\elem{i}{\Ham_{\text{dim}}}{i}$ coming from the dimerization
  Hamiltonian.}
\label{fig:1spinon-full}
\end{figure}

As we have seen, the effet of $\Ham_{\text{dim}}$ is to favor the
creation of domains with spinons at the edges. If one considers an
open chain with an odd number of sites, thus having one spinon, the
effective model for the spinon is given by
$\widetilde{\Ham}_{\text{RMG}} + \lambda
\widetilde{\Ham}_{\text{dim}}$. The comparison of the variational
approach with DMRG calculation is given on
Fig.~\ref{fig:1spinon-full}.  The agreement is pretty good and we
notice that because the confinement potential near its minimum
behaves, to zero order approximation, almost linearly with the
distance, one could expect that the tails of the wave-function are
Airy function $\propto e^{-\frac 2 3 ((i-i_0)/\xi)^{3/2}}$ rather than
pure exponential, with $\xi$ a localization length. Fitting with a
pure exponential gives a slightly worse fit (this would correspond to
straight lines on this log plot).

\subsection{Energy of a long dimer state}

We denote $\ket{[2i-1,2j]}=\ket{\MGstateLongEx}$ the singlet product
state with a long dimer between sites $2i-1$ and $2j$ ($i\leqslant
j$). This state can be seen as a singlet state between two localized
spinons. So, away from MG line ($\lambda>0$), it may have a lower
energy than the MG state. Let us apply the Hamiltonian on this state :
\begin{align*}
\Ham\ket{[2i-1,2j]}=&\frac{1}{4}\left(-3\sum_{n=1}^{i-1} \alpha_{2n-1} + \sum_{n=1}^{i-2} (\alpha_{2n}-\beta_{2n}-\beta_{2n+1})\right) \ket{[2i-1,2j]} \\ 
              +&\frac{1}{2}\sum_{n=1}^{i-2} (\alpha_{2n}-\beta_{2n}-\beta_{2n+1}) \ket{\LongDimerExcitedLeft} \\
              +&\frac{1}{4}\left(-3\sum_{n=i}^{j-1} \alpha_{2n} + \sum_{n=i+1}^{j-1} (\alpha_{2n-1}-\beta_{2n-1}-\beta_{2n})\right) \ket{[2i-1,2j]} \\
              +&\frac{1}{2}\sum_{n=i+1}^{j-1} (\alpha_{2n-1}-\beta_{2n-1}-\beta_{2n}) \ket{\LongDimerExcitedCenter} \\
              +&\frac{1}{4}\left(-3\sum_{n=j+1}^{\frac{L}{2}} \alpha_{2n-1} + \sum_{n=j+1}^{\frac{L}{2}-1} (\alpha_{2n}-\beta_{2n}-\beta_{2n+1})\right) \ket{[2i-1,2j]} \\ 
              +&\frac{1}{2}\sum_{n=j+1}^{\frac{L}{2}-1}(\alpha_{2n}-\beta_{2n}-\beta_{2n+1})\ket{\LongDimerExcitedRight}\\
              +&\frac{1}{4}(\alpha_{2i-2}+\alpha_{2i-1}-\beta_{2i-2}-\beta_{2i}+\alpha_{2i-1}+\alpha_{2j}-\beta_{2i-1}-\beta_{2j+1}) \ket{[2i-1,2j]} \\
              +&\frac{1}{2}\left[(\alpha_{2i-2}-\beta_{2i-2})\ket{[2i-3,2j]} +(\alpha_{2i-1}-\beta_{2i})\ket{[2i+1,2j]}\right] \\
              +&\frac{1}{2}\left[(\alpha_{2j-1}-\beta_{2j-1})\ket{[2i-1,2j-2]} + (\alpha_{2j}-\beta_{2j+1})\ket{[2i-1,2j+2]}\right] \\
              +&\frac{1}{4}(\beta_{2i-1}+\beta_{2j})\ket{[2i-1,2j]}\\
              +&\frac{1}{2}\left[\beta_{2i-1}\ket{\LongDimerRoofLeft} + \beta_{2j}\ket{\LongDimerRoofRight}\right]\;.
\end{align*}
Of course, this state is not exactly an eigenstate but using
\begin{equation}
\braket{[2i-1,2j]}{[2i-1\pm2,2j]}=\braket{[2i-1,2j]}{[2i-1,2j\pm2]}=-\frac{1}{2},
\end{equation}
one can calculate its energy :
\begin{equation}
\elem{[2i-1,2j]}{\Ham}{[2i-1,2j]}=-\frac{3}{4}\Big(\sum_{n=1}^{i-1}\alpha_{2n-1}+\sum_{n=i}^{j-1}\alpha_{2n}+\sum_{n=j+1}^{L/2}\alpha_{2n-1}\Big)\;,
\end{equation}
and compare it with the energy of the MG state :
\begin{align}
\elem{[2i-1,2j]}{\Ham}{[2i-1,2j]}-\elem{\text{MG}}{\Ham}{\text{MG}}&=\frac{3}{4} \Big(\sum_{n=i}^{j}\alpha_{2n-1}-\sum_{n=i}^{j-1}\alpha_{2n}\Big)\\
&=\frac{3}{4} \big(\beta_{2i-1}+\beta_{2j}\big)+\frac 3 4 \lambda\Big(\eta_{2j-1} + \sum_{n=i}^{j-1}\eta_{2n-1}-\eta_{2n}\Big)\;.
\end{align}

\newpage
\section{On the convergence of DMRG calculations}

DMRG calculations were performed using the finite-size algorithm,
targeting one or two states (for instance to determine the singlet
gap) and keeping typically from 400 to 1000 kept states. As MG are
products of dimers, they have a simple matrix-product form which makes
DMRG pretty efficient. The energies are converged to high precision.
In the presence of localized spinons and in the $S^z=0$ sector, the
local magnetization should be zero everywhere. Yet, DMRG builds up a
variational states with non-zero magnetization at the place of
localized spinons. Indeed, due to the localization of spinons, triplet
and singlet states are degenerate within an energy gap that is tiny
(we observed gaps below $10^{-8}\beta$) and controlled by the residual
magnetic couplings between spinons. Thus, the effective couplings
between spinons become so tiny that it is extremely hard for DMRG to
differentiate between the singlet or $S^z=0$ triplet state and gives a
superposition of these states as an output, with a finite local
magnetization. Still, the spinon localization and magnetization
profiles in the $S^z=1$ are very well converged.

\section{RSRG equations for the dimerized chain}
\label{app:RSRG}

Due to frustration, ferromagnetic couplings can be generated during
the RSRG scheme so one has to take into account the possibility to
generate spins higher than 1/2. The renormalized couplings, which are
here written in the general form $J_{ij}$, depend on the spin size
$s_i$. We have the following two equations corresponding to the
decimation scheme sketched in Fig.~\ref{fig:RSRGequations}:
\begin{figure}[!hb]
\centering
\includegraphics[width=0.4\columnwidth,clip]{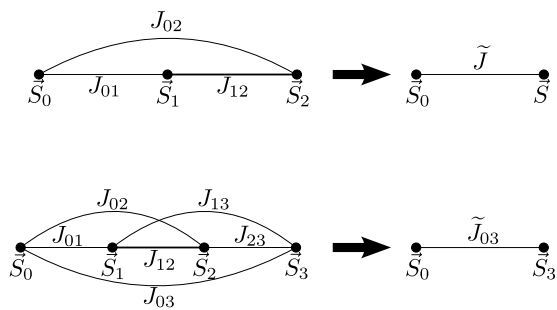}
\caption{Decimation scheme for the RSRG procedure of Model~\eqref{eq:hamiltonian}.}
\label{fig:RSRGequations}
\end{figure}
\begin{itemize}
\item if $s_1\neq s_2$ or $J_{12}<0$, we take $s=s_1+s_2$ for $J_{12}<0$ and $s=|s_1-s_2|$ for $J_{12}>0$ and we have
\begin{equation}
\widetilde{J}=\frac{s(s+1)+s_1(s_1+1)-s_2(s_2+1)}{2s(s+1)}J_{01}+\frac{s(s+1)+s_2(s_2+1)-s_1(s_1+1)}{2s(s+1)}J_{02}
\end{equation}
\item if $s_1=s_2$ and $J_{12}>0$, we have
\begin{equation}
\widetilde{J}_{03}=J_{03}+\frac{2}{3}s_1(s_1+1)\frac{(J_{01}-J_{02})(J_{23}-J_{13})}{J_{12}}
\end{equation}
\end{itemize}
In particular for the first decimations, if $\alpha_i$ is the
strongest coupling, spins $i$ and $i+1$ are decimated and the
renormalized couplings between remaining spins are :
\begin{align}
\widetilde{J}_{i-1,i+2}&=\frac{(\alpha_{i-1}-\beta_i)(\alpha_{i+1}-\beta_{i+1})}{2\alpha_i}
=\frac{(\beta_{i-1}+\lambda\eta_{i-1})(\beta_{i+2}+\lambda\eta_{i+1})}{2\alpha_i}\\
\widetilde{J}_{i-2,i+2}&=\frac{\beta_{i-1}(\alpha_{i+1}-\beta_{i+1})}{2\alpha_i}
=\frac{\beta_{i-1}(\beta_{i+2}+\lambda\eta_{i+1})}{2\alpha_i}\\
\widetilde{J}_{i-1,i+3}&=\frac{(\alpha_{i-1}-\beta_i)\beta_{i+2}}{2\alpha_i}
=\frac{(\beta_{i-1}+\lambda\eta_{i-1})\beta_{i+2}}{2\alpha_i}\\
\widetilde{J}_{i-2,i+3}&=\frac{\beta_{i-1}\beta_{i+2}}{2\alpha_i}
\end{align}
Thus for $\lambda=0$, we end up with four degenerated couplings,
implicitly reminiscent of the degeneracy of the MG domains at the RMG
point, which makes the continuation of the RSRG procedure unstable
numerically and ill-posed.

We have studied the behavior of the RSRG equations which lead to a
large-spin Griffith phase but the details will be published elsewhere.


\begin{thebibliography}{10}
\bibitem{footnote} More precisely, it is important to stress that the
  binding of spinons, lifting the degeneracy between singlet and
  triplet states, is essentially due to the confinement interaction
  between spinons see eg Refs.~\onlinecite{DimerizedChain}. At the RMG
  point, this features is preserved since MG domains remain
  degenerate. Whatever the distance between the two spinons, the
  singlet and triplet states remain degenerate.
\bibitem{footnote2}
A non-trivial exponent is expected and a specific statistical analysis
should be carried out to provide correct results.
\end{thebibliography}
\end{document}